\documentclass[fleqn,a4paper,10pt,twocolumn]{scrartcl}
\usepackage{graphicx}
\usepackage[footnotesize]{caption2}

\newcommand{\dderiv}[3][]{\frac{\mathrm{d}^{#1}{#2}}{\mathrm{d}{#3}^{#1}}}
\newcommand{\Ec}{E_\mathrm{C}}
\newcommand{\kb}{k_\mathrm{B}}

\title{A fast, primary Coulomb blockade thermometer}
\author{Tobias Bergsten, Tord Claeson and Per Delsing\\
\small{\textit{Department of Microelectronics and Nanoscience}}\\
\small{\textit{Chalmers University of Technology and G\"oteborg University}}\\
\small{\textit{SE-412 96 G\"oteborg, Su\`ede}}}
\date{}

\begin{document}

\twocolumn[
\maketitle
\vspace{-10mm}

ä\begin{abstract}
\small\quotation\centering
\begin{minipage}{142mm}
We have measured the third derivative of the current-voltage
characteristics, $\mathrm{d}^3I/\mathrm{d}V^3$, in a two-dimensional
array of small tunnel junctions using a lock-in amplifier.  We show
that this derivative is zero at a voltage which scales linearly with
the temperature and depends only on the temperature and natural
constants, thus providing a primary thermometer.  We demonstrate a
measurement method which extracts the zero crossing voltage directly
using a feedback circuit.  This method requires only one voltage
measurement, which makes it substantially faster than the original
Coulomb blockade thermometry method.
\end{minipage}\\
ä\end{abstract}
\vspace{10mm}
] 

Coulomb blockade thermometry (CBT) \cite{pekolaprl,farhang,bergstenjap}
is a primary thermometry method which is suitable for cryogenic
temperatures in the range 20\,mK - 30\,K. It is based on the
properties of the Coulomb blockade \cite{averin,grabert} in one- or
two-dimensional arrays of tunnel junctions at temperatures where the
charging energy $\Ec<\kb T$.  Here $\Ec=e^2/2C_\mathrm{eff}$, where
$C_\mathrm{eff}$ is the effective capacitance of the tunnel junctions
\cite{bergstenjap}.  In the CBT method, the first derivative of the
current-voltage characteristics (IV-curve) is measured and from the
properties of this curve the temperature can be extracted, using only
natural constants and a calculable prefactor.  The major advantage of
the CBT method is the simple electrical measurement and the
insensitivity to magnetic field. \cite{pekola_mag}

In this letter we present an alternative approach which has the
advantage of measurement speed.  Using the same type of tunnel
junction array we can measure the third derivative of the IV-curve. 
The third derivative has a zero crossing at a voltage which is (to the
first order in $\Ec/\kb T$) proportional to the temperature.  Thus,
only one measurement point, the zero crossing, is needed to measure
the temperature.  With the original method, a full
$\mathrm{d}I/\mathrm{d}V$ vs $V$ curve must be
measured.\footnote{There is also a secondary measurement mode where
only the zero-bias conductance is measured.  This requires calibration
of the tunnel junction resistance and capacitance.}

Taking the expression for $\mathrm{d}I/\mathrm{d}V$ \cite{pekolaprl}
and deriving twice with respect to $V$ the third derivative
$\mathrm{d}^3I/\mathrm{d}V^3$ can be written, to the first order in
$\Ec/\kb T$

\begin{equation}\label{eq:d3IdV3}
    \dderiv[3]{I}{V}=-\frac{Me}{R_\mathrm{T}C_\mathrm{eff}}\left(
    \frac{e}{N\kb T}\right)^3 g''\left(\frac{eV}{N\kb T}\right).
\end{equation}

Here $N$ and $M$ are the number of tunnel junctions in series and in
parallel (in the case of a two-dimensional array) respectively,
$R_\mathrm{T}$ is the tunnelling resistance of one junction at
voltages well above the Coulomb blockade and $g(x)$ is defined by
Pekola \textit{et al.} \cite{pekolaprl}, and can be written

\begin{equation}
    g(x)=\frac{(x/2)\coth(x/2)-1}{2\sinh^2(x/2)}
\end{equation}

The function $g''(x)$ becomes

\begin{equation}\label{eq:g2}
    g''(x)=\frac{\left((x/2)\coth(x/2)-1\right)
    \left(3\coth^2(x/2)-2\right)}{2\sinh^2(x/2)}
\end{equation}

Eq.  (\ref{eq:d3IdV3}) is valid in the limit $\Ec\ll\kb T$ and
$R_\mathrm{T}\gg R_\mathrm{K}=h/e^2\approx$25.8\,k$\Omega$.  Lower
temperatures and lower tunnelling resistances cause deviations which
can be calculated theoretically \cite{farhang,farhang2}.  In this
paper we have taken into account the effects due to low temperature.  The
deviations due to low resistance were not considered here, but are
estimated to be less than 1\% in the measurements presented here.

We can calculate the voltage $V_0$ at the zero-crossing of Eq.
(\ref{eq:d3IdV3}) numerically and the result is

\begin{equation}\label{eq:V0_1}
    eV_0=\pm 2.144 N\kb T.
\end{equation}

At low temperatures (where $\kb T$ approaches $\Ec$) higher order
corrections should be included \cite{farhang,kinaret}.  This gives a
correction to the zero crossing in Eq.  (\ref{eq:d3IdV3}), which is
independent of temperature:

\begin{equation}\label{eq:V0_2}
    eV_0=2.144 N\kb T - 0.465N\Ec.
\end{equation}

The sample we have measured was a two-dimensional array of $256\times
256$ tunnel junctions and each junction had an effective capacitance
of 2.2\,fF and a tunnelling resistance of 17\,k$\Omega$.  The array
was fabricated using standard shadow evaporation \cite{niemeyer,dolan} of
aluminium and in situ oxidation.  The measurements were carried out by
applying a DC voltage and an additional AC excitation (123\,Hz) to the
sample in series with a resistor $R_\mathrm{b}$ with a resistance of
20\,k$\Omega$.  The voltage over the resistor was measured with a
Stanford SRS830 lock-in amplifier which locked to the third harmonic
(369\,Hz) of the excitation AC voltage.  This signal $\delta
V_{3\omega}$ is proportional to the third derivative of the IV-curve. 
The excitation voltage $\delta V_\omega$ over the sample was measured
with another lock-in amplifier which locked to the basic frequency
(123\,Hz) and the DC voltage was measured with a voltmeter. 
Additional \mbox{low-,} high- and band-pass filtering was used to
improve the measurement.

We used a pumped $^4$He cryostat equipped with a vacuum regulator
which kept the bath at a constant pressure, and therefore constant
temperature, during the measurements.

To find the relation between the third derivative and $\delta
V_{3\omega}$ we can make a Taylor expansion of the IV-curve to the
third order and arrive at the formula

\begin{equation}\label{eq:d3I_2}
    \dderiv[3]{I}{V}=
    \frac{24}{R_\mathrm{b}}\frac{\delta V_{3\omega}}{\delta V_\omega^3},
\end{equation}

where $\delta V_{3\omega}$ and $\delta V_\omega$ are voltage
amplitudes (not rms values).

This method requires a relatively large excitation amplitude,
comparable to $V_0$, which introduces errors due to higher
derivatives.  Including fifth and seventh orders in the Taylor series
we can write

\begin{equation}\label{eq:d3I_higher}
    \frac{24}{R_\mathrm{b}}\frac{\delta V_{3\omega}}{\delta V_\omega^3}=
    \dderiv[3]{I}{V}+\frac{\delta V_\omega^2}{16}\dderiv[5]{I}{V}+
    \frac{\delta V_\omega^4}{640}\dderiv[7]{I}{V}.
\end{equation}

Knowing the amplitude of the excitation $\delta V_\omega$, the errors
due to the higher order terms can be calculated and be corrected
for.  However, right at the zero crossing of the third derivative the higher
order terms are quite small and therefore a relatively large $\delta
V_\omega$ can be used without introducing large errors.

\begin{figure}
\center
\includegraphics[width=\columnwidth]{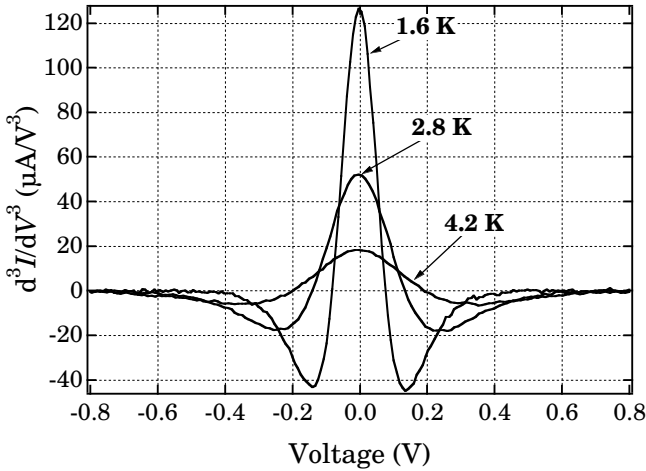}
\caption{The third derivative of the IV-curve,
$\mathrm{d}^3I/\mathrm{d}V^3$ as a function of the voltage, measured
at three different temperatures.  The voltage $V_0$ at the
zero-crossing scales linearly with the temperature. The sample
was a two-dimensional array with $256\times 256$ tunnel junctions.}
\label{gr:d3IdV3}
\end{figure}

\begin{figure}
\center
\includegraphics[width=\columnwidth]{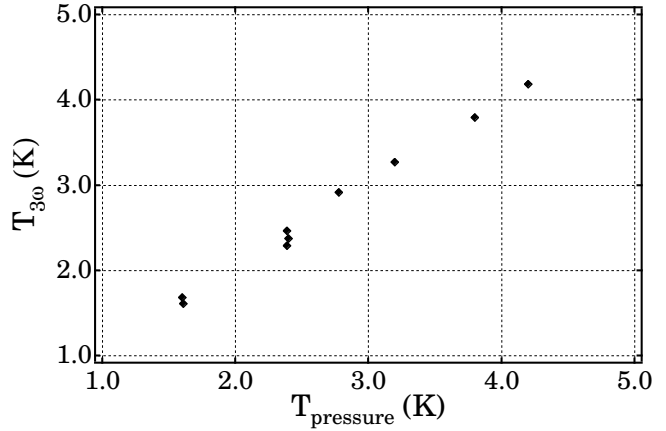}
\caption{The temperature calculated from the zero crossing (Eq. 
\ref{eq:V0_2}) versus the temperature determined from the $^4$He bath
pressure.}
\label{gr:T_vs_T}
\end{figure}

Fig. \ref{gr:d3IdV3} shows a measurement of
$\mathrm{d}^3I/\mathrm{d}V^3$ at three different temperatures.  The shape
of the curves follow the expected $g''(x)$ behaviour, and the zero
crossing follows Eq.  \ref{eq:V0_2} within a few percent.  Fig. 
\ref{gr:T_vs_T} shows the temperature calculated from the zero crossing
plotted against the temperature calculated from the $^4$He vapour
pressure \cite{its90} from 1.6\,K to 4.2\,K.

\begin{figure}
\center
\includegraphics[width=\columnwidth]{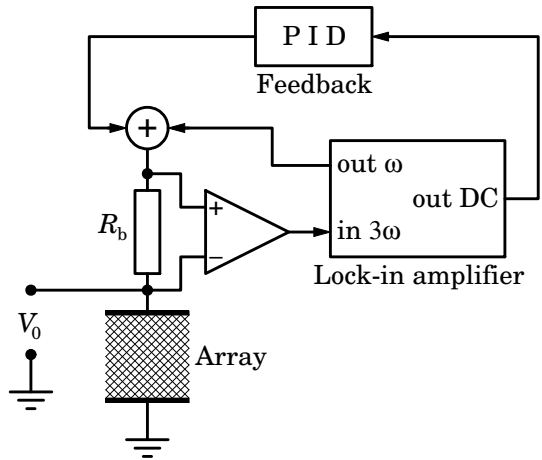}\vspace{-10mm}
\caption{The feedback loop used to demonstrate our temperature
measurement method.  The lock-in amplifier generates a sine wave which
is added to a DC voltage produced by the PID circuit.  This voltage is
applied to the sensor in series with a resistor, and the voltage over
the resistor is sensed by the lock-in amplifier.  The lock-in
amplifier is set up to detect the third harmonic of the output
frequency, to effectively measure the third derivative of the
IV-curve.  The DC output voltage is proportional to the amplitude of
the detected signal and serves as an input error signal to the PID
circuit.  With the proper feedback parameters the voltage over the
sensor array stabilises at $V_0$, as defined in Eq.  \ref{eq:V0_2}.}
\label{gr:feedback}
\end{figure}

To test the principal advantage of this measuring method compared to
the one used by Pekola \textit{et al.} we set up a feedback loop,
illustrated in Fig.  \ref{gr:feedback}.  We used a DC output voltage
from the lock-in amplifier, which was proportional to the $\delta
V_{3\omega}$ signal amplitude, as an error signal to a PID
regulator.\footnote{A PID regulator (\textbf{P}roportional,
\textbf{I}ntegrating and \textbf{D}erivating) is a general feedback
circuit with three adjustable parameters, which can be used with a wide
range of applications.} The output of the regulator was used as the DC bias
voltage and added to the AC excitation provided by the lock-in
amplifier.  The voltage over a resistor in series with the array was
applied to the input of the lock-in amplifier, which was set to
extract the third harmonic of the excitation frequency.  After we
adjusted the PID parameters to proper values, the DC voltage over the
array stabilised at the voltage $V_0$, as defined by Eq. 
\ref{eq:V0_2}.

\begin{figure}
\center
\includegraphics[width=\columnwidth]{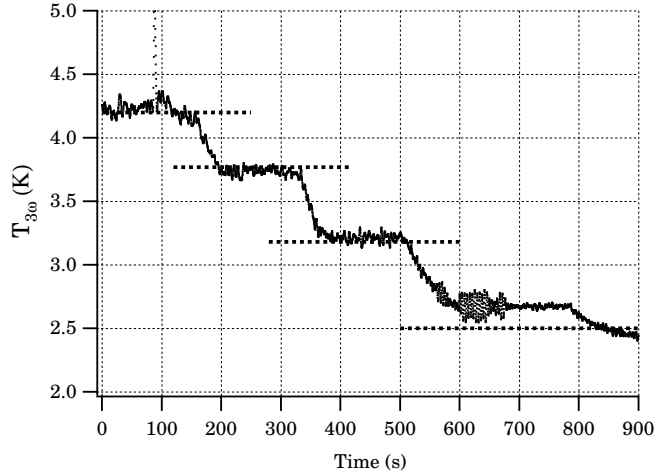}
\caption{A time trace of the temperature calculated from the
zero-crossing voltage $V_0$ (Eq.  \ref{eq:V0_2}) while the temperature
was varied in four steps.  The dashed horisontal lines represent the
temperature calculated from the vapour pressure of $^4$He
\cite{its90}.  The agreement is reasonable, except at the lowest
temperature where higher order effects and higher derivatives affect
the measurement.}
\label{gr:Td3Ivstime}
\end{figure}

As a demonstration that the voltage follows the temperature as expected, we
took a time trace of this voltage while we adjusted the temperature in
steps, by changing the bath pressure, and the result is the graph in Fig. 
\ref{gr:Td3Ivstime}, where the voltage $V_0$ is converted to
temperature using Eq.  \ref{eq:V0_2}.  The temperature steps are
evident in the figure, and agrees well with the temperature calculated
from the $^4$He vapour pressure, except for the lowest step.  The
disagreement at this step is probably due to higher derivatives and
higher order corrections to Eq.  \ref{eq:V0_2}.  \cite{farhang} At the
beginning of the fourth step, at 600 s in Fig.  \ref{gr:Td3Ivstime},
the $P$ gain of the feedback was too large and the signal started to
oscillate, but after reducing the gain (at around 680\,s in the graph)
the signal was stable again.  Note that the relatively slow time
response is not due to the thermometer or the measurement, but rather
due to the time it takes to pump down the pressure in the $^4$He bath.

While Fig.  \ref{gr:Td3Ivstime} shows that this method is working, the
precision is not very impressive in this first experiment, with
fluctuations up to 10\%.  However, there are several ways to improve the
method.  The use of a lock-in amplifier picks up the very weak third
harmonic signal below the main excitation signal and the noise, but by
notching out the basic frequency before the input we can increase the
dynamic range of the amplifier, and get a cleaner signal.  The PID
parameters can also be better optimised to the measurement, and
continuously adjusted.

Looking at Eq.  \ref{eq:d3IdV3}, it is obvious that we can make
another improvement by increasing $M$, \textit{i.e.} the number of
parallel junctions in the array.  The signal amplitude increases
linearly with $M$.  This speaks in favour of using 2D arrays with this
measurement method.  Note that we do not gain anything by decreasing
$N$, the number of junctions in series, because in order to avoid
higher order derivatives to affect the measurement we need to decrease
the excitation amplitude $\delta V_\omega$ by the same amount.  Even
though it was not done in this experiment, it would be natural to let
$\delta V_\omega$ be proportional to the temperature, to compensate
for the strong signal dependence on temperature ($\sim T^{-3}$).

In conclusion, we have measured the third derivative of the IV-curve
of a two-dimensional array of tunnel junctions.  We show,
theoretically and experimentally, that the zero crossing of this curve
scales linearly with temperature, to the first order, and provides a
primary temperature measurement.  We also demonstrate the use of a
feedback loop to create a fast primary thermometer.  The feedback loop
is possible because only one measurement is needed to get at a
quantity which is proportional to the temperature.

The Swedish Nanometer Laboratory was used to fabricate 
the samples.  This work was supported financially by the Swedish 
foundation for strategic research (SSF), by the European Union under 
the TMR program and by the Swedish research council for engineering
sciences (TFR). We would also like to thank Jari Kinaret for fruitful
discussions.


 \normalsize

\end{document}